# Comment: Demystifying Double Robustness: A Comparison of Alternative Strategies for Estimating a Population Mean from Incomplete Data

**Anastasios A. Tsiatis and Marie Davidian**

## INTRODUCTION

We congratulate Drs. Kang and Schafer (KS henceforth) for a careful and thought-provoking contribution to the literature regarding the so-called "double robustness" property, a topic that still engenders some confusion and disagreement. The authors' approach of focusing on the simplest situation of estimation of the population mean $\mu$ of a response $y$ when $y$ is not observed on all subjects according to a missing at random (MAR) mechanism (equivalently, estimation of the mean of a potential outcome in a causal model under the assumption of no unmeasured confounders) is commendable, as the fundamental issues can be explored without the distractions of the messier notation and considerations required in more complicated settings. Indeed, as the article demonstrates, this simple setting is sufficient to highlight a number of key points.

As noted eloquently by Molenberghs (2005), in regard to how such missing data/causal inference problems are best addressed, two "schools" may be identified: the "likelihood-oriented" school and the "weighting-based" school. As we have emphasized previously (Davidian, Tsiatis and Leon, 2005), we prefer to view inference from the vantage point of semiparametric theory, focusing on the assumptions embedded in the statistical models leading to different "types" of estimators (i.e., "likelihood-oriented" or "weighting-based") rather than on the forms of the estimators themselves. In this discussion, we hope to complement the presentation of the authors by elaborating on this point of view.

Throughout, we use the same notation as in the paper.

## SEMIPARAMETRIC THEORY PERSPECTIVE

As demonstrated by Robins, Rotnitzky and Zhao (1994) and Tsiatis (2006), exploiting the relationship between so-called *influence functions* and estimators is a fruitful approach to studying and contrasting the (large-sample) properties of estimators for parameters of interest in a statistical model. We remind the reader that a *statistical model* is a class of densities that could have generated the observed data. Our presentation here is for scalar parameters such as $\mu$, but generalizes readily to vector-valued parameters. If one restricts attention to estimators that are *regular* (i.e., not "pathological"; see Davidian, Tsiatis and Leon, 2005, page 263 and Tsiatis 2006, pages 26–27), then, for a parameter $\mu$ in a parametric or semiparametric statistical model, an estimator $\widehat{\mu}$ for $\mu$ based on independent and identically distributed observed data $z_i$, $i = 1, \ldots, n$, is said to be *asymptotically linear* if it satisfies

$$(1) \qquad n^{1/2}(\widehat{\mu} - \mu_0) = n^{-1/2} \sum_{i=1}^{n} \varphi(z_i) + o_p(1)$$

for $\varphi(z)$ with $E\{\varphi(z)\} = 0$ and $E\{\varphi^2(z)\} < \infty$, where $\mu_0$ is the true value of $\mu$ generating the data, and expectation is with respect to the true distribution of $z$. The function $\varphi(z)$ is the *influence function* of the estimator $\widehat{\mu}$. A regular, asymptotically linear estimator with influence function $\varphi(z)$ is consistent and asymptotically normal with asymptotic variance $E\{\varphi^2(z)\}$. Thus, there is an inextricable connection between estimators and influence functions in that the asymptotic behavior of an estimator is fully determined by its influence function, so that it suffices to focus on the influence function when discussing an estimator's properties. Many of the estimators discussed by KS are regular and asymptotically linear; in the sequel, we refer to regular and asymptotically linear estimators as simply "estimators."

*Anastasios A. Tsiatis is Drexel Professor of Statistics at North Carolina State University, Raleigh, North Carolina 27695-8203, USA e-mail: tsiatis@stat.ncsu.edu. Marie Davidian is William Neal Reynolds Professor of Statistics at North Carolina State University, Raleigh, North Carolina 27695-8203, USA e-mail: davidian@stat.ncsu.edu.*

We capitalize on this connection by considering the problem of estimating $\mu$ in the setting in KS in terms of statistical models that may be assumed for the observed data, from which influence functions corresponding to estimators valid under the assumed models may be derived. In the situation studied by KS, the "full" data that would ideally be observed are $(t, x, y)$; however, as $y$ is unobserved for some subjects, the observed data available for analysis are $z = (t, x, ty)$. As noted by KS, the MAR assumption states that $y$ and $t$ are conditionally independent given $x$; for example, $P(t=1|y,x) = P(t=1|x)$. Under this assumption, all joint densities for the observed data have the form

$$(2) \qquad p(z) = p(y|x)^{I(t=1)} p(t|x) p(x),$$

where $p(y|x)$ is the density of $y$ given $x$, $p(t|x)$ is the density of $t$ given $x$, and $p(x)$ is the marginal density of $x$. Let $p_0(z)$ be the density in the class of densities of form (2) generating the observed data (the true joint density).

One may posit different statistical models by making different assumptions on the components of (2). We focus on three such models:

I. Make no assumptions on the forms of $p(x)$ or $p(t|x)$, leaving these entirely unspecified. Make a specific assumption on $p(y|x)$, namely, that $E(y|x) = m(x, \beta)$ for some given function $m(x, \beta)$ depending on parameters $\beta$ ($p \times 1$). Denote the class of densities satisfying these assumptions as $\mathcal{M}_I$.
II. Make no assumptions on the forms of $p(x)$ or $p(y|x)$. Make a specific assumption on $p(t|x)$ that $P(t=1|x) = E(t|x) = \pi(x, \alpha)$ for some given function $\pi(x, \alpha)$ depending on parameters $\alpha$ ($s \times 1$). Here, we also require the assumption that $P(t=1|x) \geq \varepsilon > 0$ for all $x$ and some $\varepsilon$. Denote the class of densities satisfying these assumptions as $\mathcal{M}_{II}$.
III. Make no assumptions on the form of $p(x)$, but make specific assumptions on $p(y|x)$ and $p(t|x)$, namely, that $E(y|x) = m(x, \beta)$ and $P(t=1|x) = E(t|x) = \pi(x, \alpha) \geq \varepsilon > 0$ for all $x$ and some $\varepsilon$ for given functions $m(x, \beta)$ and $\pi(x, \alpha)$ depending on parameters $\beta$ and $\alpha$. The class of densities satisfying these assumptions is $\mathcal{M}_I \cap \mathcal{M}_{II}$.

All of I–III are semiparametric statistical models in that some aspects of $p(z)$ are left unspecified. Denote by $m_0(x)$ the true function $E(y|x)$ and by $\pi_0(x)$ the true function $P(t=1|x) = E(t|x)$ corresponding to the true density $p_0(z)$.

Semiparametric theory yields the form of all influence functions corresponding to estimators for $\mu$ under each of the statistical models I–III. As discussed in Tsiatis (2006, page 52), loosely speaking, a consistent and asymptotically normal estimator for $\mu$ in a statistical model has the property that, for all $p(z)$ in the class of densities defined by the model, $n^{1/2}(\widehat{\mu} - \mu) \xrightarrow{\mathcal{D}(p)} \mathcal{N}\{0, \sigma^2(p)\}$, where $\xrightarrow{\mathcal{D}(p)}$ means convergence in distribution under the density $p(z)$, and $\sigma^2(p)$ is the asymptotic variance of $\widehat{\mu}$ under $p(z)$.

If model I is correct, then $m_0(x) = m(x, \beta)$ for some $\beta$, and it may be shown (e.g., Tsiatis, 2006, Section 4.5) that all estimators for $\mu$ have influence functions of the form

$$(3) \qquad m_0(x) - \mu + ta(x)\{y - m_0(x)\}$$

for arbitrary functions $a(x)$ of $x$. If model II is correct, then $\pi_0(x) = \pi(x, \alpha)$ for some $\alpha$, and all estimators for $\mu$ have influence functions of the form

$$(4) \qquad \frac{ty}{\pi_0(x)} + \frac{t - \pi_0(x)}{\pi_0(x)} h(x) - \mu$$

for arbitrary $h(x)$, which is well known from Robins, Rotnitzky and Zhao (1994). If model III is correct, then $m_0(x) = m(x, \beta)$ and $\pi_0(x) = \pi(x, \alpha)$ for some $\beta$ and $\alpha$, and influence functions for estimators $\widehat{\mu}$ have the form

$$(5) \qquad \begin{aligned} & m_0(x) - \mu + ta(x)\{y - m_0(x)\} \\ & + \frac{t - \pi_0(x)}{\pi_0(x)} h(x) \end{aligned}$$

for arbitrary $a(x)$ and $h(x)$. Depending on forms of $m(x, \beta)$ as a function of $\beta$ and $\pi(x, \alpha)$ as a function of $\alpha$, there will be restrictions on the forms of $a(x)$ and $h(x)$; see below.

We now consider estimators discussed by KS from the perspective of influence functions. The regression estimator $\widehat{\mu}_{OLS}$ in (7) of KS comes about naturally if one assumes model I is correct. In terms of influence functions, $\widehat{\mu}_{OLS}$ may be motivated by considering the influence function (3) with $a(x) = 0$, as this leads to the estimator $n^{-1} \sum_{i=1}^{n} m(x_i, \beta)$. In fact, although KS do not discuss it, the "imputation estimator" $\widehat{\mu}_{IMP} = n^{-1} \sum_{i=1}^{n} \{t_i y_i + (1 - t_i) m(x_i, \beta)\}$ may be motivated by taking $a(x) = 1$ in (3). Of course, in practice, $\beta$ must be estimated. In general, (3) implies that all estimators for $\mu$ that are consistent and asymptotically normal if model I is correct must be asymptotically equivalent to an estimator of the form

$$(6) \quad n^{-1} \sum_{i=1}^{n} [m(x_i, \widehat{\beta}) + t_i \widetilde{a}(x_i) \{y_i - m(x_i, \widehat{\beta})\}],$$

where $\beta$ is estimated by solving an estimating equation $\sum_{i=1}^{n} t_i A(x_i, \beta) \{y_i - m(x_i, \beta)\} = 0$ for $A(x, \beta)$ ($p \times 1$). Because $\beta$ is estimated, the influence function of the estimator (6) with a particular $\widetilde{a}(x)$ will not be exactly equal to (3) with $a(x) = \widetilde{a}(x)$; instead, it may be shown that the influence function of (6) is of form (3) with $a(x)$ in (3) equal to

$$(7) \quad \begin{aligned} &\widetilde{a}(x) - E[\{\pi_0(x)\widetilde{a}(x) - 1\} m_\beta^T(x, \beta_0)] \\ &\quad \cdot [E\{\pi_0(x) A(x, \beta_0) m_\beta^T(x, \beta_0)\}]^{-1} \\ &\quad \cdot A(x, \beta_0), \end{aligned}$$

where $m_\beta(x, \beta)$ is the vector of partial derivatives of elements of $m(x, \beta)$ with respect to $\beta$, and $\beta_0$ is such that $m_0(x) = m(x, \beta_0)$.

The IPW estimator $\widehat{\mu}_{IPW\text{-}POP}$ in (3) of KS and its variants arise if one assumes model II. In particular, $\widehat{\mu}_{IPW\text{-}POP}$ can be motivated via the influence function (4) with $h(x) = -\mu$. The estimator $\widehat{\mu}_{IPW\text{-}NR}$ in (4) of KS follows from (4) with $h(x) = -E[y\{1 - \pi(x)\}]/E[\{1 - \pi(x)\}]$. In fact, if one restricts $h(x)$ in (4) to be a constant, then, using the fact that the expectation of the square of (4) is the asymptotic variance of the estimator, one may find the "best" such constant minimizing the variance as $h(x) = -E[y\{1 - \pi(x)\}/\pi(x)]/E[\{1 - \pi(x)\}/\pi(x)]$. An estimator based on this idea was given in (10) of Lunceford and Davidian (2004, page 2943). In general, as for model I, (4) implies that all estimators for $\mu$ that are consistent and asymptotically normal if model II is correct must be asymptotically equivalent to an estimator of the form

$$(8) \quad n^{-1} \sum_{i=1}^{n} \left\{ \frac{t_i y_i}{\pi(x_i, \widehat{\alpha})} + \frac{t_i - \pi(x_i, \widehat{\alpha})}{\pi(x_i, \widehat{\alpha})} \widetilde{h}(x_i) \right\},$$

where $\widehat{\alpha}$ is estimated by solving an equation of the form $\sum_{i=1}^{n} \{t_i - \pi(x_i, \alpha)\} B(x_i, \alpha) = 0$ for some ($s \times$ 1) $B(x_i, \alpha)$, almost always maximum likelihood for binary regression. As above, because $\alpha$ is estimated, the influence function of (8) is equal to (4) with $h(x)$ equal to

$$(9) \quad \begin{aligned} &\widetilde{h}(x) - E[\pi_\alpha^T(x, \alpha_0) \{m_0(x) + \widetilde{h}(x)\}/\pi_0(x)] \\ &\quad \cdot [E\{B(x, \alpha_0) \pi_\alpha^T(x, \alpha_0)\}]^{-1} \\ &\quad \cdot B(x, \alpha_0) \pi_0(x), \end{aligned}$$

where $\pi_\alpha(x, \alpha)$ is the vector of partial derivatives of elements of $\pi(x, \alpha)$ with respect to $\alpha$, and $\alpha_0$ satisfies $\pi_0(x) = \pi(x, \alpha_0)$.

Doubly robust (DR) estimators are estimators that are consistent and asymptotically normal for models in $\mathcal{M}_I \cup \mathcal{M}_{II}$, that is, under the assumptions of model I or model II. When the true density $p_0(z) \in \mathcal{M}_I \cap \mathcal{M}_{II}$, then the influence function of any such DR estimator must be equal to (3) with $a(x) = 1/\pi_0(x)$ or, equivalently, equal to (4) with $h(x) = -m_0(x)$. Accordingly, when $p_0(z) \in \mathcal{M}_I \cap \mathcal{M}_{II}$, that is, both models have been specified correctly, all such DR estimators will have the same asymptotic variance. This also implies that, if both models are correctly specified, the asymptotic properties of the estimator do not depend on the methods used to estimate $\beta$ and $\alpha$.

KS discuss strategies for constructing DR estimators, and they present several specific examples: $\widehat{\mu}_{BC\text{-}OLS}$ in their equation (8); the estimators below (8) using POP or NR weights, which we denote as $\widehat{\mu}_{BC\text{-}POP}$ and $\widehat{\mu}_{BC\text{-}NR}$, respectively; the estimator $\widehat{\mu}_{WLS}$ in their equation (10); $\widehat{\mu}_{\pi\text{-}cov}$ in their equation (12); and a version of $\widehat{\mu}_{\pi\text{-}cov}$ equal to the estimator proposed by Scharfstein, Rotnitzky and Robins (1999) and Bang and Robins (2005), which we denote as $\widehat{\mu}_{SRR}$. The results for these estimators under the "Correct-Correct" scenarios ($\mathcal{M}_I \cap \mathcal{M}_{II}$) in Tables 5–8 of KS are consistent with the asymptotic properties above. We note that $\widehat{\mu}_{\pi\text{-}cov}$ is not DR under $\mathcal{M}_I \cup \mathcal{M}_{II}$ because of the additional assumption that the mean of $y$ given $\pi$ must be equal to a linear combination of basis functions in $\pi$. Making this additional assumption may not be unreasonable in practice; however, strictly speaking, it takes $\widehat{\mu}_{\pi\text{-}cov}$ outside the class of DR estimators discussed here, and hence we do not consider it in the remainder of this section. However, $\widehat{\mu}_{SRR}$ is still in this class.

KS suggest that a characteristic distinguishing the performance of DR estimators is whether or not the estimator is within or outside the augmented inverse-probability weighted (AIPW) class. We find this distinction artificial, as all of the above estimators $\widehat{\mu}_{BC\text{-}OLS}$, $\widehat{\mu}_{BC\text{-}POP}$, $\widehat{\mu}_{BC\text{-}NR}$, $\widehat{\mu}_{WLS}$ and $\widehat{\mu}_{SRR}$ can be expressed in an AIPW form. Namely, all of these estimators are algebraically exactly of the form (8) with $\widetilde{h}(x_i)$ replaced by a term $-\widehat{\gamma} - m(x_i, \widehat{\beta})$, where $\widehat{\gamma}_{BC\text{-}OLS} = \widehat{\gamma}_{WLS} = \widehat{\gamma}_{SRR} = 0$,

$$\widehat{\gamma}_{BC\text{-}POP}$$
$$(10) \quad = \frac{n^{-1} \sum_{i=1}^{n} (t_i/\widehat{\pi}_i)(y_i - \widehat{m}_i)}{n^{-1} \sum_{i=1}^{n} t_i/\widehat{\pi}_i} \quad \text{and}$$
$$\widehat{\gamma}_{BC\text{-}NR}$$
$$= \frac{n^{-1} \sum_{i=1}^{n} (t_i(1-\widehat{\pi}_i)/\widehat{\pi}_i)(y_i - \widehat{m}_i)}{n^{-1} \sum_{i=1}^{n} t_i(1-\widehat{\pi}_i)/\widehat{\pi}_i},$$

where we write $\widehat{\pi}_i = \pi(x_i, \widehat{\alpha})$ and $\widehat{m}_i = m(x_i, \widehat{\beta})$ for brevity. For $\widehat{\mu}_{WLS}$ and $\widehat{\mu}_{SRR}$, this identity follows from the fact that $\sum_{i=1}^{n} \frac{t_i}{\widehat{\pi}_i}(y_i - \widehat{m}_i) = 0$, which for $\widehat{\mu}_{WLS}$ holds because KS restrict to $m(x, \beta) = x^T \beta$, with $x$ including a constant term. Thus, we contend that issues of performance under $\mathcal{M}_I \cup \mathcal{M}_{II}$ are not linked to whether or not a DR estimator is AIPW, but, rather, are a consequence of forms of the influence functions of estimators under $\mathcal{M}_I$ or $\mathcal{M}_{II}$. In particular, under model II, it follows that the above estimators have influence functions of the form (4) with $h(x)$ equal to (9) with $\widetilde{h}(x) = -\{\gamma^* + m(x, \beta^*)\}$, where $\gamma^*$ and $\beta^*$ are the limits in probability of $\widehat{\gamma}$ and $\widehat{\beta}$, respectively. Thus, features determining performance of these estimators when model II is correct are how close $\gamma^* + m(x, \beta^*)$ is to $m_0(x)$ and how $\alpha$ is estimated, where maximum likelihood is the optimal choice. In fact, this perspective reveals that, for fixed $m(x, \beta)$, using ideas similar to those in Tan (2006), the optimal choice of $\widehat{\gamma}$ is as in $\widehat{\gamma}_{BC\text{-}NR}$ with $t_i(1 - \widehat{\pi}_i)/\widehat{\pi}_i$ replaced by $t_i(1 - \widehat{\pi}_i)/\widehat{\pi}_i^2$.

Similarly, under model I, the influence functions of these estimators are of the form (3) with $a(x)$ equal to (7) with $\widetilde{a}(x) = \psi_1/\pi(x, \alpha^*) + \psi_2$, where $\alpha^*$ is the limit in probability of $\widehat{\alpha}$ and $\psi_1 = 1$ and $\psi_2 = 0$ for $\widehat{\mu}_{BC\text{-}OLS}$, $\widehat{\mu}_{WLS}$ and $\widehat{\mu}_{SRR}$; $\psi_1 = 1/E\{\pi_0(x)/\pi(x, \alpha^*)\}$ and $\psi_2 = 0$ for $\widehat{\mu}_{BC\text{-}POP}$; and $\psi_1$ and $\psi_2$ for $\widehat{\mu}_{BC\text{-}NR}$ are more complicated expectations involving $\pi_0(x)$ and $\pi(x, \alpha^*)$. Thus, under model I, features determining performance of these estimators are the form of $\widetilde{a}(x)$ and how $\beta$ is estimated through the choice of $A(x, \beta)$.

We may interpret some of the results in Tables 5, 6 and 8 of KS in light of these observations. Under the "$\pi$-model Correct–$y$-model Incorrect" scenario ($\mathcal{M}_{II} \cap \mathcal{M}_I^c$), $\widehat{\mu}_{BC\text{-}OLS}$, $\widehat{\mu}_{WLS}$ and $\widehat{\mu}_{SRR}$ show some nontrivial differences in performance, which, from above, are likely attributable to differences in $m(x, \beta^*)$. Under the "$\pi$-model Incorrect–$y$-model Correct" ($\mathcal{M}_I \cap \mathcal{M}_{II}^c$), all three estimators share the same $\widetilde{a}(x)$ but use different methods to estimate $\beta$, so that any differences are dictated entirely by the choice of $A(x, \beta)$. The poor performance of $\widehat{\mu}_{SRR}$ can be understood from this perspective: "$\beta$" for this estimator is actually $\beta$ in the model $m(x, \beta)$ used by the other two estimators concatenated by an additional element, the coefficient of $\widehat{\pi}_i^{-1}$. The $A(x, \beta)$ for $\widehat{\mu}_{SRR}$ thus involves a design matrix that is unstable for small $\widehat{\pi}_i$, consistent with the comment of KS at the end of their Section 3.

In summary, we believe that studying the performance of estimators via their influence functions can provide useful insights. Our preceding remarks refer to large-sample performance, which depends directly on the influence function. Estimators with the same influence function can exhibit different finite-sample properties. It may be possible via higher-order expansions to gain an understanding of some of this behavior; to the best of our knowledge, this is an open question.

## BOTH MODELS INCORRECT

The developments in the previous section are relevant in $\mathcal{M}_I \cup \mathcal{M}_{II}$. Key themes of KS are performance of DR and other estimators outside this class; that is, when both the models $\pi(x, \alpha)$ and $m(x, \beta)$ are incorrectly specified, and choice of estimator under these circumstances.

One way to study performance in this situation is through simulation. KS have devised a very interesting and instructive "specific simulation scenario that highlights some important features of various estimators. In particular, the KS scenario emphasizes the difficulties encountered with some of the DR estimators when $\pi(x_i, \widehat{\alpha})$ is small for some $x_i$. Indeed, in our experience, poor performance of DR and IPW estimators in practice can result from few small $\pi(x_i, \widehat{\alpha})$. When there are small $\pi(x_i, \widehat{\alpha})$, as noted KS, responses are not observed for some portion of the $x$ space. Consequently, estimators like $\widehat{\mu}_{OLS}$ rely on extrapolation into that part of the $x$ space. KS have constructed a scenario where failure to observe $y$ in a portion of the $x$ space can wreak havoc on some estimators that make use of the $\pi(x_i, \widehat{\alpha})$ but has minimal impact on the quality of extrapolations for these $x$ based on $m(x, \widehat{\beta})$.

One could equally well build a scenario where the $x$ for which $y$ is unobserved are highly influential for the regression $m(x,\beta)$ and hence could result in deleterious performance of $\widehat{\mu}_{OLS}$. We thus reiterate the remark of KS that, although simulations can be illuminating, they cannot yield broadly applicable conclusions.

Given this, we offer some thoughts on other strategies for deriving estimators that may have some robustness properties under the foregoing conditions, that is, offer good performance outside $\mathcal{M}_I \cup \mathcal{M}_{II}$. One approach may be to search outside the class of DR estimators valid under $\mathcal{M}_I \cup \mathcal{M}_{II}$. For example, as suggested by the simulations of KS, estimators in the spirit of $\widehat{\mu}_{\pi\text{-}cov}$, which impose additional assumptions rendering them DR in the strict sense only in a subset of $\mathcal{M}_I \cup \mathcal{M}_{II}$, may compensate for this restriction by yielding more robust performance outside $\mathcal{M}_I \cup \mathcal{M}_{II}$; further study along these lines would be interesting. An alternative tactic for searching outside $\mathcal{M}_I \cup \mathcal{M}_{II}$ may be to consider the form of influence functions (5) for estimators valid under $\mathcal{M}_I \cap \mathcal{M}_{II}$. For instance, a "hybrid" estimator of the form

$$n^{-1} \sum_{i=1}^{n} \bigg[ m(x_i, \widehat{\beta}) I\{\pi(x_i, \widehat{\alpha}) < \delta\}$$
$$+ \bigg\{ \frac{t_i y_i}{\pi(x_i, \widehat{\alpha})} + \frac{t_i - \pi(x_i, \widehat{\alpha})}{\pi(x_i, \widehat{\alpha})} \widetilde{h}(x_i) \bigg\}$$
$$\cdot I\{\pi(x_i, \widehat{\alpha}) \geq \delta\} \bigg],$$

for $\delta$ small, may take advantage of the desirable properties of both $\widehat{\mu}_{OLS}$ and DR estimators.

A second possible strategy for identifying robust estimators arises from the following observation. Consider the estimator

$$(11) \quad n^{-1} \sum_{i=1}^{n} \bigg\{ \frac{t_i y_i}{\pi(x_i)} - \frac{t_i - \pi(x_i)}{\pi(x_i)} m(x_i, \widehat{\beta}) \bigg\}.$$

If $\pi(x_i) = \pi(x_i, \widehat{\alpha})$, then (11) yields one form of a DR estimator. If $\pi(x_i) \equiv 1$, then (11) results in the imputation estimator. If $\pi(x_i) = \infty$, (11) reduces to $\widehat{\mu}_{OLS}$. This suggests that it may be possible to develop estimators based on alternative choices of $\pi(x_i)$ that may have good robustness properties. For example, a method for obtaining estimators $\pi(x_i, \widehat{\alpha})$ that shrinks these toward a common value may prove fruitful. The suggestion of KS to move away from logistic regression models for $\pi(x_i, \alpha)$ is in a similar spirit.

Finally, we note that yet another approach to developing estimators would be to start with the premise that one make no parametric assumption on the forms of $E(y|x)$ and $E(t|x)$ beyond some mild smoothness conditions. Here, it is likely that first-order asymptotic theory, as in the previous section, may no longer be applicable. It may be necessary to use higher-order asymptotic theory to make progress in this direction; see, for example, Robins and van der Vaart (2006).

## CONCLUDING REMARKS

We again compliment the authors for their thoughtful and insightful article, and we appreciate the opportunity to offer our perspectives on this important problem. We look forward to new methodological developments that may overcome some of the challenges brought into focus by KS in their article.

## ACKNOWLEDGMENT

This research was supported in part by Grants R01-CA051962, R01-CA085848 and R37-AI031789 from the National Institutes of Health.

## REFERENCES

Bang, H. and Robins, J. M. (2005). Doubly robust estimation in missing data and causal inference models. *Biometrics* **61** 962–972. MR2216189

Davidian, M., Tsiatis, A. A. and Leon, S. (2005). Semiparametric estimation of treatment effect in a pretest-posttest study without missing data. *Statist. Sci.* **20** 261–301. MR2189002

Lunceford, J. K. and Davidian, M. (2004). Stratification and weighting via the propensity score in estimation of causal treatment effects: A comparative study. *Statistics in Medicine* **23** 2937–2960.

Molenberghs, G. (2005). Discussion of "Semiparametric estimation of treatment effect in a pretest–posttest study with missing data," by M. Davidian, A. A. Tsiatis and S. Leon. *Statist. Sci.* **20** 289–292. MR2189002

Robins, J. M., Rotnitzky, A. and Zhao, L. P. (1994). Estimation of regression coefficients when some regressors are not always observed. *J. Amer. Statist. Assoc.* **89** 846–866. MR1294730

Robins, J. and van der Vaart, A. (2006). Adaptive nonparametric confidence sets. *Ann. Statist.* **34** 229–253. MR2275241

Scharfstein, D. O., Rotnitzky, A. and Robins, J. M. (1999). Rejoinder to "Adjusting for nonignorable dropout using semiparametric nonresponse models." *J. Amer. Statist. Assoc.* **94** 1135–1146. MR1731478

Tan, Z. (2006). A distributional approach for causal inference using propensity scores. *J. Amer. Statist. Assoc.* **101** 1619–1637. MR2279484

Tsiatis, A. A. (2006). *Semiparametric Theory and Missing Data*. Springer, New York. MR2233926